\documentclass[conference]{IEEEtran}
\usepackage{cite}
\usepackage{graphicx}
\usepackage{amsmath,amssymb}
\usepackage{array,booktabs}
\usepackage{url}
\usepackage{xcolor}
\usepackage[hidelinks]{hyperref}
\hypersetup{pdftitle={Uncertainty-Guided UAV Spectrum Cartography with Deep-Unfolded Online Tensor Decomposition},pdfauthor={Shangjie Zhuang, Jiahui Liang, Shijian Gao}}
\graphicspath{{figures/}}

\begin{document}
\title{Uncertainty-Guided UAV Spectrum Cartography with Deep-Unfolded Online Tensor Decomposition}
\author{\IEEEauthorblockN{Shangjie Zhuang, Jiahui Liang, and Shijian Gao}
\IEEEauthorblockA{Internet of Things Thrust, The Hong Kong University of Science and Technology (Guangzhou), Guangzhou 511400, China\\
Email: \{szhuang962, jliang097\}@connect.hkust-gz.edu.cn, shijiangao@hkust-gz.edu.cn}}
\maketitle
\begin{abstract}
Spectrum cartography is crucial for spectrum-aware resource management in low-altitude networks, where uncrewed aerial vehicles (UAVs) collect spectrum measurements to reconstruct power spectral density (PSD) maps. However, limited energy and sensing bandwidth make measurements sparse in space and incomplete in frequency. To collect these measurements efficiently, the UAV actively plans its next move based on the latest map and its uncertainty, which requires rapid online reconstruction. We therefore propose an active online spectrum cartography framework. We first develop online deep-unfolded tensor decomposition (ODU-TD) for rapid map updates. An ensemble of reconstructors then estimates uncertainty to select informative sensing targets, and a learning-based policy determines the UAV movement and sensing bandwidth under the energy budget. Experiments show that ODU-TD achieves an approximately 27-fold speedup over online tensor decomposition and the lowest normalized mean square error (NMSE) among the compared reconstructors under sparse spatial and spectral observations, and the proposed framework reduces the NMSE by at least 55\% compared with the representative baselines.
\end{abstract}

\section{Introduction}
Spectrum cartography aims to reconstruct power spectral density (PSD) radio maps, which describe the received power jointly across space and frequency, from sparse spectrum measurements~\cite{yang2026survey}. Such maps provide environmental awareness, which is crucial for spectrum-related resource management in low-altitude networks~\cite{romero2022radio}.

Tensor-based models couple spatial propagation fields with corresponding spectra and analyze map recoverability~\cite{zhang2020spectrum}. Interpolation has been further integrated to accommodate irregular spatial sampling and incomplete frequency sampling~\cite{sun2024iibtd}. Autoencoders and generative adversarial networks exploit frequency-spatial correlations to infer maps at unmeasured frequencies~\cite{zhou2023frequencyspatial}. Further studies combine neural models with structured tensor factorization~\cite{shrestha2022deep}, recover spectrum maps from incomplete and corrupted observations via tensor completion~\cite{wang2025remtensor}, and integrate compressive sensing with wideband spectrum mapping~\cite{shen2023compressed}. Taken together, these works show how structural models and learned representations exploit spatial-spectral correlations to reconstruct PSD maps from incomplete measurements. However, these methods focus on offline reconstruction from fixed measurement sets, whereas the measurements are collected by energy-limited uncrewed aerial vehicles (UAVs) in low-altitude airspace. 

In contrast, online approaches incrementally update the map as new measurements are collected during the flight, which makes it possible to exploit the current reconstruction for decision making. For example, an active sensing approach uses map uncertainty to guide UAVs toward informative measurement locations, thereby improving sampling efficiency~\cite{shrestha2023spectrum}. Gaussian process ensembles and travel-cost-aware active learning select informative locations while limiting UAV travel~\cite{polyzos2024bayesian}. Under limited onboard energy, UAVs have been guided to high-value measurement locations for aerial channel knowledge map construction~\cite{chen2026energyckm}, and deep reinforcement learning has been used to optimize their trajectories for radio map updating~\cite{li2025atdqn}. Bayesian uncertainty and graph-based reinforcement learning have also been combined to plan informative and energy-efficient trajectories~\cite{lu2025uram}. These studies demonstrate the value of adaptive spatial sampling for radio map construction. However, they mainly target signal-strength radio maps and decide only where to measure, whereas in PSD mapping the limited sensing bandwidth of the UAV allows only a subset of frequency bands to be observed at each location. Deciding which and how many bands to sense under the energy budget thus relies on the PSD map and its uncertainty updated after every measurement, which calls for rapid online PSD reconstruction.

\begin{figure}[!t]
    \centering
    \includegraphics[width=0.88\columnwidth]{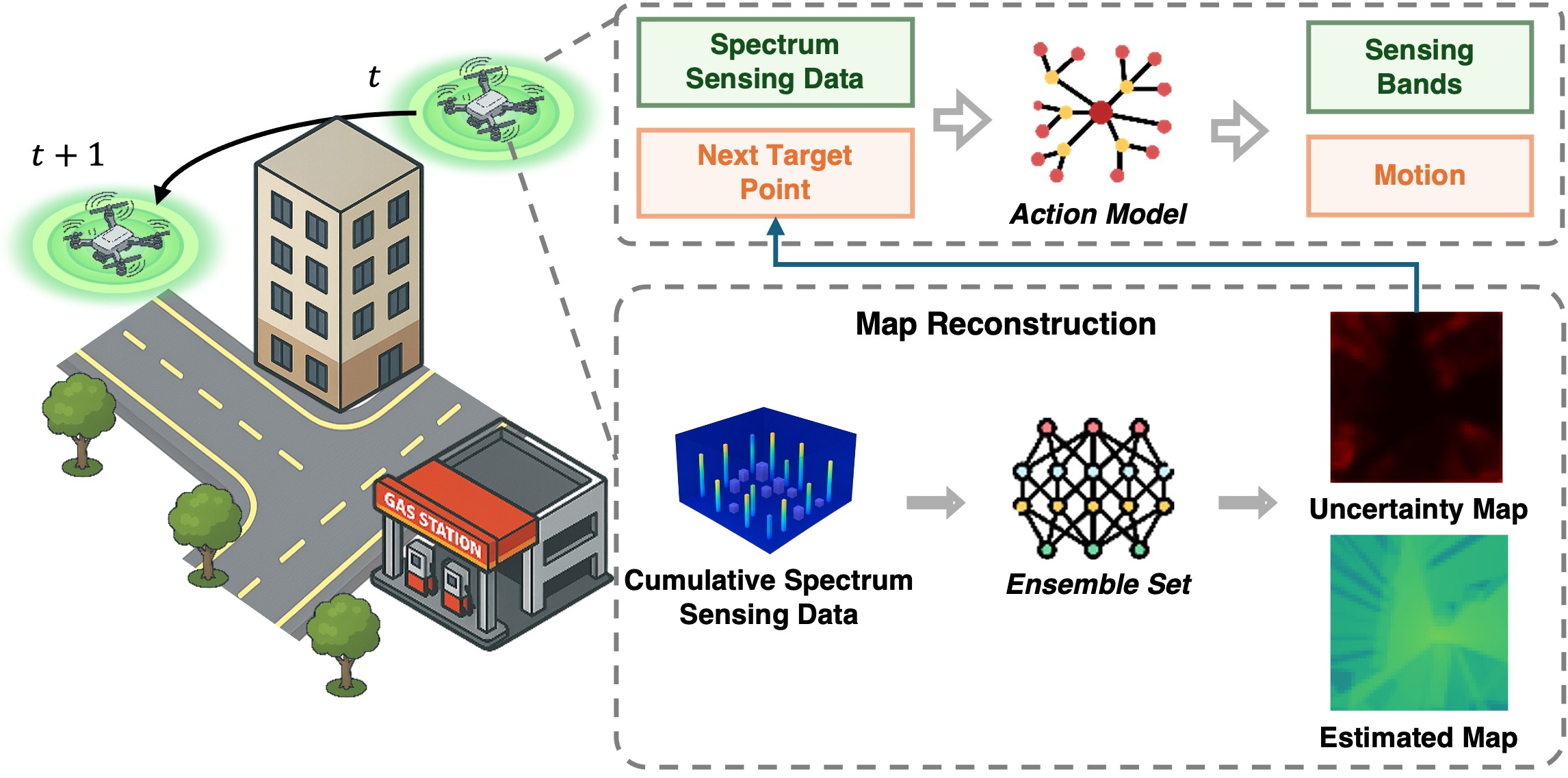}
    \caption{Uncertainty-guided closed loop of the proposed active online spectrum cartography framework.}
    \label{fig:system_overview}
\end{figure}

In this paper, we propose an active online spectrum cartography framework, as illustrated in Fig.~\ref{fig:system_overview}. At each step, an online deep-unfolded tensor decomposition (ODU-TD) reconstructor rapidly updates the PSD map from cumulative measurements. Based on the updated map, an ensemble of reconstructors estimates uncertainty to identify an informative target location and frequency band, and a learned policy selects the UAV movement and the number of consecutive bands to sense under the remaining energy budget. The new measurement is then added, and the cycle repeats. The main contributions are summarized as follows:
\begin{itemize}
\item We develop an online deep-unfolded tensor decomposition method, which replaces the computational bottleneck with a learnable module and unfolds alternating updates into fixed stages for timely and accurate updates under partial observations.
\item We propose an active online spectrum cartography framework that uses ensemble uncertainty to select informative sampling targets and a learned policy to select the movement and sensing bandwidth under energy constraints.
\item Experiments show that ODU-TD accelerates online tensor decomposition and achieves the lowest error under partial observations, and that the framework achieves lower reconstruction error than representative baselines.
\end{itemize}

\section{System Model and Problem Formulation}
\subsection{PSD Radio Map Model}
We discretize the horizontal $x$--$y$ plane into $N_x\times N_y$ grid cells indexed by $\mathcal{G}=\{(i,j)\mid 1\leq i \leq N_x,1\leq j \leq N_y\}$. Each grid cell is represented by its center at coordinates $(i,j)$ in grid units. The monitored spectrum is divided into $K$ frequency bands indexed by $k=1,\ldots,K$. The PSD radio map is represented by a tensor $\boldsymbol{\mathcal{H}}\in\mathbb{R}_{+}^{N_x\times N_y\times K}$, where $\boldsymbol{\mathcal{H}}(i,j,k)$ denotes the received PSD at grid cell $(i,j)$ and frequency band $k$. The radio environment is generated by $R$ emitters. Let $\mathbf{S}_r\in\mathbb{R}^{N_x\times N_y}$ and $\boldsymbol{\phi}_r=[\phi_1^{r},\ldots,\phi_K^{r}]^{\top}$ denote the large-scale spatial propagation field and power spectrum of source $r$, respectively. Following the tensor decomposition (TD) model that exploits spatial and spectral correlations~\cite{zhang2020spectrum,sun2024iibtd}, the PSD map is
\begin{equation}
    \boldsymbol{\mathcal{H}}
    =
    \sum_{r=1}^{R} \mathbf{S}_r\circ\boldsymbol{\phi}_r.
    \label{eq:btd_map_model}
\end{equation}

\subsection{Partial Spectrum Observation Model}
We model the online spectrum cartography process over $T$ time slots. At slot $t$, the UAV grid coordinate is $\boldsymbol{u}_t\in\mathcal{G}$. The UAV senses $q_t$ consecutive frequency bands centered on the target frequency band. Let $\boldsymbol{\omega}_t\in\{0,1\}^{K}$ denote the corresponding binary mask, where $[\boldsymbol{\omega}_t]_k=1$ if frequency band $k$ is selected and zero otherwise. Hence, $\sum_{k=1}^{K}[\boldsymbol{\omega}_t]_k=q_t$. Measurements are affected by small-scale fading $\eta_{r,t}^{(k)}\sim\mathcal{N}(0,\sigma_\eta^2)$ and receiver noise $\epsilon_t^{(k)} \sim \mathcal{N}(0,\sigma_\epsilon^2)$, both independent across frequency bands. The aggregate observation disturbance is
$n_t^{(k)}=\sum_{r=1}^{R}\phi_k^{r}\eta_{r,t}^{(k)}+\epsilon_t^{(k)}$, and $\boldsymbol{n}_t=[n_t^{(1)},\ldots,n_t^{(K)}]^{\top}$. The sparse wideband scanning vector is
\begin{equation}
    \boldsymbol{y}_t
    =
    \boldsymbol{\omega}_t\odot
    \left[\boldsymbol{\mathcal{H}}(\boldsymbol{u}_t,:)+\boldsymbol{n}_t\right],
    \label{eq:partial_observation}
\end{equation}
where $\boldsymbol{\mathcal{H}}(\boldsymbol{u}_t,:)=[\boldsymbol{\mathcal{H}}(\boldsymbol{u}_t,1),\ldots,\boldsymbol{\mathcal{H}}(\boldsymbol{u}_t,K)]^{\top}$ is the full PSD vector at the current UAV position.

\subsection{Energy Consumption Model}
The energy consumption model includes both movement and sensing costs. The movement energy cost is
\begin{equation}
E_t^{\rm mov}=E_{\rm fly}\|\boldsymbol{u}_t-\boldsymbol{u}_{t-1}\|_1
+E_{\rm hov}\mathbb{I}\{\boldsymbol{u}_t=\boldsymbol{u}_{t-1}\},
\end{equation}
where $E_{\rm fly}$ and $E_{\rm hov}$ denote the flying and hovering energy costs, respectively. The sensing energy cost is
\begin{equation}
E_t^{\rm sen}=E_{\rm sen}\frac{q_t\log_2(1+q_t)}{q_{\max}\log_2(1+q_{\max})},
\label{eq:sensing_energy_cost}
\end{equation}
where $q_{\max}$ denotes the maximum number of sensed bands; the sensing cost increases with $q_t$. The total energy consumption at slot $t$ is $E_t=E_t^{\rm mov}+E_t^{\rm sen}$, and the energy budget is $E_{\max}$.

\subsection{Problem Formulation}
At slot $t$, policy $\pi$ selects the action $a_t=(m_t^{\rm u},q_t)$, where $m_t^{\rm u}$ is a feasible UAV movement and $q_t\in\mathcal{Q}$ is the number of sensed bands. After $T$ slots, the normalized mean square error (NMSE) is $\mathcal{E}_T=\|\widehat{\boldsymbol{\mathcal{H}}}_T-\boldsymbol{\mathcal{H}}\|_F^2/{\|\boldsymbol{\mathcal{H}}\|_F^2}.$ The active online spectrum cartography problem can be formulated as a constrained optimization problem $\mathbf{P0}$:
\begin{subequations}
\begin{align}
\min_{\pi} \quad 
&\mathbb{E}_{\pi}\{\mathcal{E}_T\}
\label{eq:conference_objective}\\
\mathrm{s.t.}\quad 
&q_t\in\mathcal{Q},
\label{eq:conference_sensing}\\
&\sum_{t=1}^{T}E_t\leq E_{\max}.
\label{eq:conference_energy}
\end{align}
\end{subequations}

\section{Online Spectrum Cartography}
\label{sec:odu_td}

Online spectrum cartography requires rapid and accurate map updates under partial observations to support uncertainty-guided sensing. This section first reviews the offline TD method and then extends it to sequential measurements for online reconstruction. A deep-unfolded version is subsequently developed to accelerate the updates and improve the reconstruction accuracy.

\subsection{Recap of the Tensor Decomposition}

Given a fixed set $\mathcal{D}$ of $M$ measurements, each consisting of a location $\boldsymbol{z}_m$ and the corresponding PSD values, offline TD~\cite{sun2024iibtd} fits local propagation models to jointly estimate the power spectra and spatial propagation fields in Eq.~\eqref{eq:btd_map_model}. For source $r$, offline TD approximates the propagation field around the center of grid cell $(i,j)$ using the local polynomial $f_{ij}^{r}(\boldsymbol{z}) = \boldsymbol{x}_{ij}^{\top}(\boldsymbol{z})\boldsymbol{\theta}_{ij}^{r},$ where $\boldsymbol{x}_{ij}(\boldsymbol{z})$ contains $D_{\rm p}$ polynomial features centered at $(i,j)$. The corresponding coefficient vector is $\boldsymbol{\theta}_{ij}^{r}\in\mathbb{R}^{D_{\rm p}\times 1}$. Stacking all $R$ coefficient vectors gives $\boldsymbol{\Theta}_{ij}=[(\boldsymbol{\theta}_{ij}^{1})^{\top},\ldots,(\boldsymbol{\theta}_{ij}^{R})^{\top}]^{\top}$.
The propagation estimate of source $r$ at grid cell $(i,j)$ is $f_{ij}^{r}(i,j)=\boldsymbol{e}_r^{\top}\boldsymbol{\Theta}_{ij}\approx[\mathbf{S}_r]_{ij}$. Here, $\boldsymbol{e}_r\in\mathbb{R}^{D_{\rm p}R\times 1}$ is the selection vector whose $[(r-1)D_{\rm p}+1]$-th entry is one and all other entries are zero.

The PSD measurements are recorded in $\boldsymbol{\Gamma}\in\mathbb{R}^{M\times K}$. The binary mask $\boldsymbol{\psi}\in\{0,1\}^{M\times K}$ identifies the available frequency bands, where $[\boldsymbol{\psi}]_{m,k}=1$ if frequency band $k$ is available in measurement $m$ and zero otherwise. Let $\boldsymbol{\Phi}$ collect the PSD of $R$ sources, with its $r$-th row given by $\boldsymbol{\phi}_r^{\top}\in\mathbb{R}^{1\times K}$. Define $\mathbf{X}_{ij}=[\boldsymbol{x}_{ij}(\boldsymbol{z}_1),\ldots,\boldsymbol{x}_{ij}(\boldsymbol{z}_{M})]\in\mathbb{R}^{D_{\rm p}\times M}$. The diagonal matrix $\mathbf{Q}_{ij}\in\mathbb{R}^{M\times M}$ contains the spatial kernel weights and assigns greater influence to measurements near grid cell $(i,j)$. The spatial weights and availability mask form $\mathbf{W}_{ij} = (\mathbf{I}_K\otimes\mathbf{Q}_{ij})
    \operatorname{diag}\!\left[\operatorname{vec}(\boldsymbol{\psi})\right],$
where $\mathbf{I}_K$ is the $K\times K$ identity matrix. The corresponding regression residual is $\boldsymbol{r}_{ij} (\boldsymbol{\Theta}_{ij},\boldsymbol{\Phi})  =  \operatorname{vec}(\boldsymbol{\Gamma})-(\boldsymbol{\Phi}^{\top}\otimes\mathbf{X}_{ij}^{\top})
\boldsymbol{\Theta}_{ij}.$ The offline TD reconstruction problem~$\mathbf{P1}$ is formulated as
\begin{equation}
    \begin{aligned}
        \min_{\substack{\{\boldsymbol{\Theta}_{ij}\},\boldsymbol{\Phi}, \{\mathbf{S}_r\}}}
        &\sum_{(i,j)\in\mathcal{G}}
        \left\|\mathbf{W}_{ij}
        \boldsymbol{r}_{ij}(\boldsymbol{\Theta}_{ij},\boldsymbol{\Phi})
        \right\|_2^2 \\
        &+\nu\sum_{\substack{(i,j)\in\mathcal{G}\\ r=1,\ldots,R}}
        \left(\boldsymbol{e}_r^{\top}\boldsymbol{\Theta}_{ij}
        -[\mathbf{S}_r]_{ij}\right)^2
        +\lambda\sum_{r=1}^{R}\|\mathbf{S}_r\|_* \\
        \mathrm{s.t.}\quad
        &[\boldsymbol{\Phi}]_{r,k}\geq0,\quad \forall r,k.
    \end{aligned}
    \label{eq:iibtd_objective}
\end{equation}

The first term in the objective of problem~$\mathbf{P1}$ minimizes the weighted residuals between the PSD measurements and model predictions. The second term enforces consistency between the local estimate $\boldsymbol{e}_r^{\top}\boldsymbol{\Theta}_{ij}$ and $[\mathbf{S}_r]_{ij}$. The third term applies nuclear-norm regularization to $\mathbf{S}_r$ to promote low-rank propagation fields. Offline TD solves problem~$\mathbf{P1}$ by alternating the updates of $\{\boldsymbol{\Theta}_{ij}\}$, $\boldsymbol{\Phi}$, and $\{\mathbf{S}_r\}$ until convergence, then reconstructs $\widehat{\boldsymbol{\mathcal{H}}}=\sum_{r=1}^{R}\widehat{\mathbf{S}}_r\circ\widehat{\boldsymbol{\phi}}_r$.

\subsection{Spectrum Cartography via Online TD}

At slot $t$, these definitions are applied to the cumulative measurement set $\mathcal{D}_t$. As the measurement set grows, reapplying offline TD would repeatedly process $\mathcal{D}_{t-1}$ even when only a few new samples arrive. The online TD method therefore reuses the previous estimates while retaining the block-alternating structure of Eq.~\eqref{eq:iibtd_objective}. New measurements are accumulated into a batch for each reconstruction update; the estimates remain unchanged between updates. At an update, the local coefficients are recomputed only for cells affected by measurements accumulated since the previous reconstruction.

\emph{Update of $\boldsymbol{\Theta}_{ij}$:}
Let $\mathbf{M}_t^{\rm aff}\in\{0,1\}^{N_x\times N_y}$ be the affected-cell mask, where $[\mathbf{M}_t^{\rm aff}]_{ij}=1$ if the spatial-kernel support of grid cell $(i,j)$ covers at least one UAV sample in the accumulated batch. For each cell with $[\mathbf{M}_t^{\rm aff}]_{ij}=1$, online TD updates the local coefficients as
\begin{equation}
    \begin{aligned}
        \boldsymbol{\Theta}_{ij,t}
        ={}&\underset{\boldsymbol{\Theta}_{ij}}{\arg\min}\quad
        \left\|\mathbf{W}_{ij,t}\boldsymbol{r}_{ij,t}
        (\boldsymbol{\Theta}_{ij},\boldsymbol{\Phi}_{t-1})\right\|_2^2\\[-1mm]
        &+\nu\sum_{r=1}^{R}
        \left(\boldsymbol{e}_r^{\top}\boldsymbol{\Theta}_{ij}
        -[\mathbf{S}_{r,t-1}]_{ij}\right)^2.
    \end{aligned}
    \label{eq:seq_theta_update}
\end{equation}
For cells with $[\mathbf{M}_t^{\rm aff}]_{ij}=0$, online TD retains $\boldsymbol{\Theta}_{ij,t}=\boldsymbol{\Theta}_{ij,t-1}$. Reusing these coefficients avoids redundant computation and improves the efficiency of online updates.

\emph{Update of $\boldsymbol{\Phi}$:}
Because $\boldsymbol{\Phi}$ is shared across grid cells, online TD initializes it with $\boldsymbol{\Phi}_{t-1}$ and solves
\begin{equation}
    \begin{aligned}
        \boldsymbol{\Phi}_{t}
        =\underset{\boldsymbol{\Phi}\geq\mathbf{0}}{\arg\min}\quad
        &\sum_{(i,j)\in\mathcal{G}}
        \left\|\mathbf{W}_{ij,t}
        \boldsymbol{r}_{ij,t}
        (\boldsymbol{\Theta}_{ij,t},\boldsymbol{\Phi})
        \right\|_2^2.
    \end{aligned}
    \label{eq:seq_phi_update}
\end{equation}

\emph{Update of $\mathbf{S}_r$:}
Online TD forms $\boldsymbol{\Psi}_{r,t}\in\mathbb{R}^{N_x\times N_y}$ with $[\boldsymbol{\Psi}_{r,t}]_{ij}=\boldsymbol{e}_r^{\top}\boldsymbol{\Theta}_{ij,t}$ and updates $\mathbf{S}_r$ by solving the low-rank subproblem
\begin{equation}
    \mathbf{S}_{r,t}
    =
    \underset{\mathbf{S}_r}{\arg\min}
    \nu\sum_{(i,j)\in\mathcal{G}}
    \left([\boldsymbol{\Psi}_{r,t}]_{ij}-[\mathbf{S}_r]_{ij}\right)^2
    +\lambda\|\mathbf{S}_r\|_*.
    \label{eq:iibtd_sr_problem}
\end{equation}
This subproblem is solved by iterative singular value thresholding (SVT) initialized at $\mathbf{S}_{r,t-1}$. The warm start reuses the previous solution and reduces the number of iterations required for convergence.

To make the differences in computational complexity explicit, we consider square grids with $N_x=N_y=N_{\rm g}$. Within each alternating iteration, the local coefficients $\boldsymbol{\Theta}_{ij}$ are updated by Eq.~\eqref{eq:seq_theta_update} only at affected grid cells, whereas the shared spectra $\boldsymbol{\Phi}$ are updated by Eq.~\eqref{eq:seq_phi_update} using residual information aggregated across the grid. For fixed $R$, $K$, and $D_{\rm p}$, these updates cost $\mathcal{O}(\|\mathbf{M}_t^{\rm aff}\|_0|\mathcal{D}_t|)$ and $\mathcal{O}(N_{\rm g}^2|\mathcal{D}_t|)$, respectively. In contrast, the update of $\mathbf{S}_r$ requires repeated dense singular value decompositions (SVDs). With $J_{\rm svt}$ SVT iterations, its cost is $\mathcal{O}(RJ_{\rm svt}N_{\rm g}^3)$. Hence, the updates of $\boldsymbol{\Theta}_{ij}$ and $\boldsymbol{\Phi}$ are relatively inexpensive, while repeated SVDs make the update of $\mathbf{S}_r$ the main computational bottleneck.

\subsection{Deep-Unfolded Online Spectrum Cartography}

Deep unfolding converts a number of model-based iterations into a trainable sequence of network stages while preserving the algorithmic structure~\cite{monga2021algorithm}. In radio map estimation~\cite{johnson2026factor}, it unfolds a factor-decomposed convex recovery algorithm into learnable blocks, improving reconstruction performance. Since online TD retains an alternating-update structure and its cost is dominated by the update of $\mathbf{S}_r$, we unfold the alternating updates into $T_{\rm DU}$ stages and replace iterative SVT with a learned module, yielding ODU-TD. Moreover, retaining model-based local interpolation and spectral estimation preserves explicit relationships among measurements, source spectra, and local estimates.

\begin{figure}[!t]
    \centering
    \includegraphics[width=0.8\columnwidth]{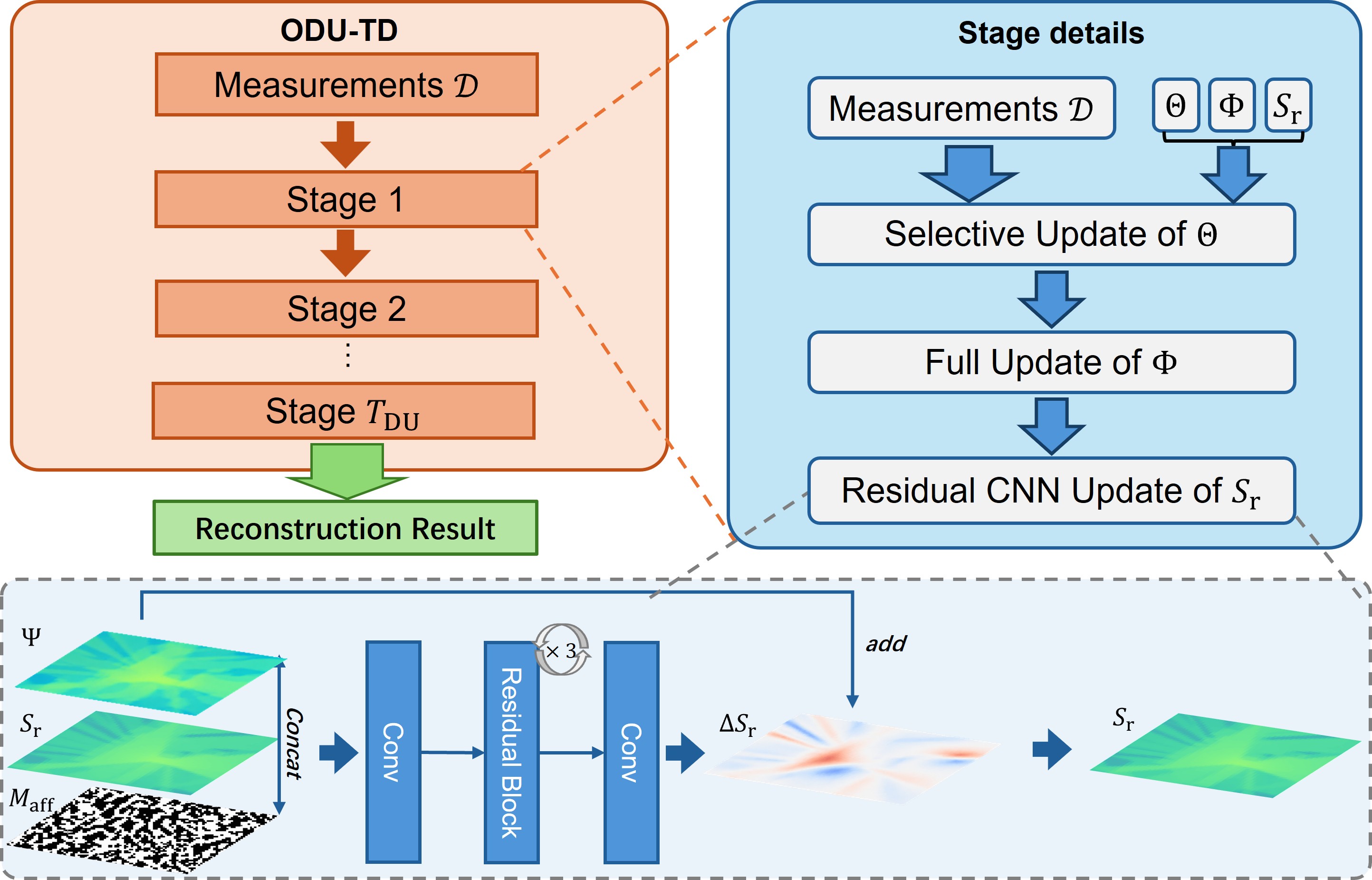}
    \caption{Structure of the proposed deep-unfolded online reconstruction network.}
    \label{fig:odutd_architecture}
\end{figure}

As illustrated in Fig.~\ref{fig:odutd_architecture}, ODU-TD uses a convolutional residual network to refine the spatial propagation fields. The network concatenates three inputs: the updated interpolation $\boldsymbol{\Psi}_r^{(\ell+1)}$, the previous-stage source map $\mathbf{S}_r^{(\ell)}$, and the affected-cell mask $\mathbf{M}_t^{\rm aff}$. The convolutions capture neighborhood-dependent propagation patterns beyond global nuclear-norm regularization. Since the spatial fields are shared across all bands in Eq.~\eqref{eq:btd_map_model}, refining them also improves the estimates at unobserved bands. This is particularly beneficial under partial observations, where the interpolation error of the propagation fields increases as fewer bands are observed~\cite[Proposition~4]{sun2024iibtd}. An input convolution fuses the inputs, residual blocks aggregate neighboring information, and an output convolution produces the residual
$\Delta\mathbf{S}_r^{(\ell+1)}=\operatorname{Prox}_{\vartheta_\ell}\{[\boldsymbol{\Psi}_r^{(\ell+1)},\mathbf{S}_r^{(\ell)},\mathbf{M}_t^{\rm aff}]\}$, where $\vartheta_\ell$ denotes the network parameters of stage $\ell$. The updated spatial propagation map is
\begin{equation}
    \begin{aligned}
        \bigl[\mathbf{S}_r^{(\ell+1)}\bigr]_{ij}
        ={}&\ln\!\left(
            1+e^{
            \bigl[\boldsymbol{\Psi}_r^{(\ell+1)}\bigr]_{ij}
            +\alpha_\ell\bigl[\Delta\mathbf{S}_r^{(\ell+1)}\bigr]_{ij}
            }
        \right),\\
        &\hspace{35mm}(i,j)\in\mathcal{G},
    \end{aligned}
    \label{eq:du_sr_update}
\end{equation}
where $\alpha_\ell$ scales the correction to the model-based estimate and the Softplus function enforces nonnegativity. After $T_{\rm DU}$ stages, the final source maps and spectra reconstruct $\widehat{\boldsymbol{\mathcal{H}}}_t$ through Eq.~\eqref{eq:btd_map_model}.

Training combines map, source-map, and observation-level supervision through
\begin{equation}
    \mathcal{L}
    =
    \mathcal{L}_{\rm err}
    +\lambda_S\mathcal{L}_S
    +\lambda_{\rm obs}\mathcal{L}_{\rm obs},
    \label{eq:du_loss}
\end{equation}
where $\lambda_S$ and $\lambda_{\rm obs}$ are nonnegative weights. The primary loss $\mathcal{L}_{\rm err}$ is defined as the NMSE between the reconstructed and ground-truth PSD maps. The auxiliary term $\mathcal{L}_S$ supervises individual source maps, directly constraining spatial refinement beyond the combined tensor error. Finally, $\mathcal{L}_{\rm obs}$ penalizes mismatch at observed entries, discouraging refinements that conflict with the supplied measurements.

The learned module refines each source map through local convolutions and pointwise operations. For a convolutional neural network with fixed depth, kernel sizes, and channel widths, these operations cost $\mathcal{O}(N_{\rm g}^2)$ per source map per stage. Refining $R$ source maps over $T_{\rm DU}$ stages therefore costs $\mathcal{O}(RT_{\rm DU}N_{\rm g}^2)$, whereas each dense SVD required by SVT costs $\mathcal{O}(N_{\rm g}^3)$. For fixed iteration and stage counts, replacing SVT with the learned module reduces source-map refinement from cubic to quadratic complexity in $N_{\rm g}$. The fixed stage count also removes convergence-dependent repetition of the alternating updates.

\section{Uncertainty-Guided Active Sampling}

Under limited energy and sensing bandwidth, the UAV should prioritize informative measurements to improve reconstruction accuracy. This section builds on the online reconstruction to form an uncertainty-guided active sampling scheme, where ensemble uncertainty selects an informative target location and frequency band, and a UAV policy determines the movement and the number of sensed bands under the energy budget.

\subsection{Uncertainty-Guided Target Selection}

Ensemble methods estimate uncertainty from the disagreement among multiple estimators. Evaluating $N_{\rm ens}$ members on the same observations yields estimates $\widehat{\boldsymbol{\mathcal{H}}}_t^{(e)}$ with weights $w_{e,t}$ satisfying $\sum_{e}w_{e,t}=1$. Given their weighted average $\bar{\boldsymbol{\mathcal{H}}}_t=\sum_{e}w_{e,t}\widehat{\boldsymbol{\mathcal{H}}}_t^{(e)}$, the resulting uncertainty is
\begin{equation}
    \boldsymbol{\mathcal{U}}_t(i,j,k)
    =
    \sum_{e=1}^{N_{\rm ens}}w_{e,t}
    \big[\widehat{\boldsymbol{\mathcal{H}}}_t^{(e)}(i,j,k)
    -\bar{\boldsymbol{\mathcal{H}}}_t(i,j,k)\big]^2.
    \label{eq:ensemble_uncertainty_map}
\end{equation}

Averaging over frequency gives the location-level uncertainty $[\mathbf{U}_t^{\rm sp}]_{ij}=K^{-1}\sum_{k=1}^{K}\boldsymbol{\mathcal{U}}_t(i,j,k)$. To identify the target frequency band, we define the frequency-aware score
\begin{equation}
    \boldsymbol{\mathcal{S}}_t^{\rm f}(i,j,k)
    =
    \boldsymbol{\mathcal{U}}_t(i,j,k)-\beta_f\boldsymbol{\mathcal{V}}_t(i,j,k),
    \label{eq:frequency_aware_score}
\end{equation}
where $\boldsymbol{\mathcal{V}}_t$ contains normalized visit counts and $\beta_f$ penalizes repeated frequency selections. The highest-scoring unobserved band at each candidate cell enters the joint target selection.

Let $\mathcal{C}_t^{\rm u}$ contain sampling-valid grid cells within the current search range that have at least one unobserved frequency band. For each candidate grid cell $(i,j)$, $\widehat{k}_t(i,j)$ denotes its unobserved band with the largest frequency-aware score in Eq.~\eqref{eq:frequency_aware_score}. The target location is selected as
\begin{equation}
    \boldsymbol{v}^\star
    =
    \arg\max_{(i,j)\in\mathcal{C}_t^{\rm u}}
    \left\{
    \lambda_u[\mathbf{U}_t^{\rm sp}]_{ij}
    +\boldsymbol{\mathcal{S}}_t^{\rm f}[i,j,\widehat{k}_t(i,j)]
    \right\}.
    \label{eq:uncertainty_target_selection}
\end{equation}
Here, $\lambda_u$ balances spatial and frequency-specific informativeness. To limit search complexity, $\mathcal{C}_t^{\rm u}$ is restricted to a local Manhattan neighborhood, which is temporarily expanded globally when the local uncertainty stops decreasing.

\subsection{Energy-Constrained Movement and Bandwidth Planning}
\label{sec:ppo_policy}

The sensing target identifies an informative location and band, while the UAV policy decides how to reach it and how many bands to sense. Sensing more bands yields more observations per slot but consumes energy needed for subsequent measurements, so these decisions are coupled over time and are learned by an actor--critic policy, whose design is detailed as follows.

\textbf{1) Observation and state:}
The actor observes the UAV status (position, remaining energy, and sensing width) and the target context (the uncertainty around $\boldsymbol{u}_t$ and at the target entry, and the direction, path distance, and band of the target). The UAV status indicates the energy available for future sensing, whereas the target context reflects how informative the target is and how costly it is to reach. The critic additionally receives the NMSE and map-update status, which are used only for value estimation during training.

\textbf{2) Action space:}
At slot $t$, $a_t=(m_t^{\rm u},q_t)$ selects a stay or four-neighbor move and $q_t\in\mathcal{Q}$ consecutive bands centered on $\widehat{k}_t(\boldsymbol{v}^\star)$. A larger $q_t$ observes more bands at the current location at a higher sensing cost. Invalid or over-budget actions are masked to satisfy constraint~\eqref{eq:conference_energy}.

\textbf{3) Reward design:}
The reconstruction reward based on NMSE is sparse, as it is available only when the map is updated. To provide intermediate feedback for UAV movement, we adopt reward shaping as in~\cite{li2025atdqn}, introduce an auxiliary progress reward, and define the shaped reward as
\begin{equation}
    r_t=\lambda_{\rm N}r_t^{\rm N}+\lambda_{\rm P}r_t^{\rm P},
    \label{eq:ppo_reward}
\end{equation}
where $\lambda_{\rm N}$ and $\lambda_{\rm P}$ balance the two components. The reconstruction reward $r_t^{\rm N}$ encourages NMSE reduction relative to the previous map update. The progress reward $r_t^{\rm P}$ rewards moving toward the sensing target and penalizes moving away. 

\textbf{4) Policy training:}
The actor and critic are trained by proximal policy optimization (PPO)~\cite{schulman2017ppo}, which clips the policy ratio to stabilize training. Generalized advantage estimation (GAE) computes advantages and return targets from the collected rewards and critic values, and the networks are updated with the clipped surrogate objective, a value loss, and an entropy bonus that encourages exploration. During policy training, the ODU-TD ensemble is pretrained and kept fixed, and only the actor is used for decision making after training. Hyperparameters are listed in Table~\ref{tab:training_settings}.

\section{Experimental Results}
\subsection{Dataset and Settings}
We construct the PSD map dataset from the large-scale propagation fields of RadioMapSeer~\cite{levie2021radiounet}, which provides building layouts and physics-based propagation maps, and ARM-Omni~\cite{gao2026farm}, which provides low-altitude aerial radio maps. These fields are cropped into $100\times100$ grid scenes. For each spatial field, we synthesize eight nonnegative power spectra $\boldsymbol{\phi}\in\mathbb{R}_{+}^{K}$ with $K=30$, each formed from squared sinc components and normalized such that $\sum_{k=1}^{K}\phi_k=K$. Pairing them with the field via Eq.~\eqref{eq:btd_map_model} yields eight dense $100\times100\times30$  PSD tensors per base scene. Base scenes are divided into training, validation, and test subsets with an $80\%/10\%/10\%$ ratio.

\begin{table}[!t]
    \centering
    \setlength{\abovecaptionskip}{2pt}
    \caption{Training and Configuration parameters.}
    \label{tab:training_settings}
    \begingroup
    \footnotesize
    \setlength{\tabcolsep}{2.0pt}
    \renewcommand{\arraystretch}{0.94}
    \begin{tabular*}{\columnwidth}{@{\extracolsep{\fill}}lc@{\hspace{8pt}}lc@{}}
        \toprule
        \multicolumn{2}{c}{ODU-TD training} & \multicolumn{2}{c}{PPO training} \\
        \cmidrule(lr){1-2}\cmidrule(lr){3-4}
        Unfolded stages & 3 & Optimizer & Adam \\
        Optimizer & AdamW & Learning rate & $10^{-4}$ \\
        Learning rate & $10^{-4}$ & Discount factor $\gamma$ & 0.99 \\
        Batch size & 2 & GAE parameter $\lambda_{\rm GAE}$ & 0.95 \\
        Max. epochs & 150 & Clip coefficient $\epsilon_{\rm PPO}$ & 0.2 \\
        & & Epochs per update & 6 \\
        \midrule
        \multicolumn{4}{c}{Configuration} \\
        \cmidrule(lr){1-4}
        Grid size & $100\times100$ & Frequency bands $K$ & 30 \\
        Grid spacing & 2 m/cell & UAV altitude & 50 m \\
        UAV step size & 4 cells & Band set $\mathcal{Q}$ & $\{4,6,8\}$ \\
        Energy budget $E_{\max}$ & 8500 J & Episode horizon $T$ & 200 \\
        $E_{\rm fly}/E_{\rm hov}$ & 12/8 J & $E_{\rm sen}$ & 5 J \\
        \bottomrule
    \end{tabular*}
    \endgroup
\end{table}
\subsection{Online Reconstruction Efficiency}

To emulate online operation, UAV measurements are fed to offline TD, online TD, and ODU-TD one at a time in the same order, and each method updates the PSD map after every new measurement. For each method, we record the runtime of each reconstruction module and the cumulative solver time over the entire sequence. Table~\ref{tab:training_settings} lists the training and configuration parameters.

\begin{table}[!t]
    \centering
    \setlength{\abovecaptionskip}{2pt}
    \caption{Mean runtime per call of reconstruction modules.}
    \label{tab:runtime_update}
    \begingroup
    \footnotesize
    \setlength{\tabcolsep}{2.5pt}
    \begin{tabular*}{\columnwidth}{@{\extracolsep{\fill}}cccc@{}}
        \toprule
        Method     & $\boldsymbol{\Theta}_{ij}$ (ms) & $\boldsymbol{\Phi}$ (ms) & $\mathbf{S}_r$ (ms) \\
        \midrule
        ODU-TD          & 1.43          & 1.19        & 0.73       
        \\
        online TD       & 1.45          & 1.07        & 183.16     
        \\
        offline TD      & 15.13         & 2.34        & 403.51     
        \\
        \bottomrule
    \end{tabular*}
    \endgroup
\end{table}

Table~\ref{tab:runtime_update} lists the mean runtime per call of each module. By updating coefficients only at affected cells, online TD reduces the $\boldsymbol{\Theta}_{ij}$ cost from $15.13$ ms to $1.45$ ms, leaving the SVT-based $\mathbf{S}_r$ update as the bottleneck at $183.16$ ms per call. ODU-TD replaces SVT with the learned module and reduces this cost to $0.73$ ms, while the $\boldsymbol{\Theta}_{ij}$ and $\boldsymbol{\Phi}$ costs remain comparable. As a result, the cumulative solver time over the measurement sequence decreases from $42.87$ s to $1.58$ s, an approximately $27$-fold speedup over online TD. This speedup enables timely map updates after each measurement, which is required for subsequent sampling decisions.

\begin{figure}[!t]
    \centering
    \includegraphics[width=\columnwidth]{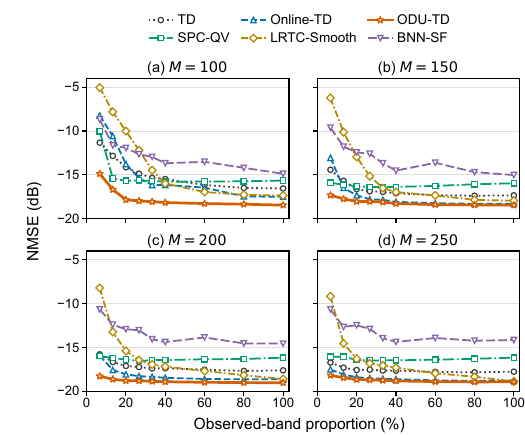}
    \caption{NMSE versus observed-band proportion for (a) $M=100$, (b) $M=150$, (c) $M=200$, and (d) $M=250$; TD and Online-TD denote offline and online TD. NMSE is averaged across scenes in linear scale.}
    \label{fig:band_count}
\end{figure}

\subsection{Reconstruction Accuracy under Partial Observations}
To evaluate the reconstruction accuracy under partial observations, ODU-TD is compared with offline and online TD, which share its signal model but refine source maps by SVT, and with three representative reconstructors: smooth PARAFAC with quadratic variation (SPC-QV)~\cite{yokota2016spc}, solved by L-BFGS; low-rank tensor completion with spatial smoothness (LRTC-Smooth)~\cite{schaeufele2019smooth}, solved by Douglas--Rachford splitting; and a spatial-frequency Bayesian neural network (BNN-SF) that reconstructs all $K$ bands from sparse measurements and their masks, following URAM~\cite{lu2025uram} with Monte Carlo dropout~\cite{gal2016dropout}.

We consider $q\in\{2,4,6,8,10,12,18,24,30\}$ observed bands out of $K=30$, with measurements collected at $M\in\{100,150,200,250\}$ spatial locations. Within each scene, all methods use the same ordered locations and frequency masks. As shown in Fig.~\ref{fig:band_count}, ODU-TD achieves the lowest NMSE in all settings, and its advantage is most evident when both spatial and spectral observations are sparse. Moreover, for $M\geq150$, its NMSE varies only slightly with the observed-band proportion, indicating that the learned spatial refinement compensates for missing spectral observations. In contrast, the baselines either rely heavily on spectral observations or saturate at higher error levels. These results confirm that ODU-TD maintains accurate reconstruction under partial observations, allowing the UAV to sense fewer bands without substantially degrading the reconstruction when sufficient locations are measured.

\subsection{Active Sampling Performance}
\begin{figure}[!t]
    \centering
    \includegraphics[width=0.9\columnwidth]{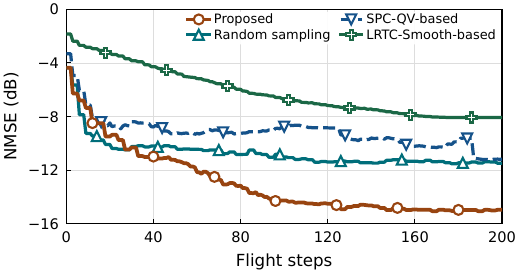}
    \caption{NMSE versus flight steps under active sampling. Proposed combines ODU-TD and PPO; Random sampling replaces PPO with a random policy, while SPC-QV-based and LRTC-Smooth-based replace ODU-TD with the corresponding reconstructors.}
    \label{fig:ppo_step_nmse}
\end{figure}

Fig.~\ref{fig:ppo_step_nmse} compares the active sampling performance, where each method collects its own observations. Random sampling reduces the NMSE faster within the first 30 steps but soon saturates at around $-11$ dB, whereas the proposed framework keeps improving as the flight proceeds and finally reduces the NMSE by 55.0\% relative to random sampling. With the same PPO policy, the proposed framework also reduces the final NMSE by 58.2\% and 79.5\% compared with the SPC-QV-based and LRTC-Smooth-based variants, respectively. These results confirm that uncertainty-guided active sampling selects informative measurements that continue to improve the reconstruction accuracy under the energy budget, and that the framework further benefits from the accurate online reconstruction of ODU-TD.

\section{Conclusions}
This paper developed an active online spectrum cartography framework under a limited UAV energy budget and incomplete PSD observations. Deep-unfolded online reconstruction enables timely map updates as measurements arrive, while ensemble-based uncertainty identifies informative sensing locations and frequency bands. Through an uncertainty-guided feedback loop, the framework determines where the UAV moves and how many bands it senses under the energy budget. Experiments show that the proposed reconstruction method effectively accelerates online map updates and maintains low reconstruction error under sparse spectral observations. In active sampling, the proposed framework achieves lower reconstruction error than the evaluated baselines.
\bibliographystyle{IEEEtran}
\bibliography{refs}
\end{document}